\documentclass[10pt]{iopart}

\usepackage[utf8]{inputenc}
\usepackage{subcaption}
\usepackage[table,xcdraw,dvipsnames]{xcolor}
\usepackage{adjustbox}
\usepackage{amssymb}
\expandafter\let\csname equation*\endcsname\relax

\expandafter\let\csname endequation*\endcsname\relax
\usepackage{amsmath}
\usepackage{upgreek}
\def \be {\begin{equation}}
\def \ee {\end{equation}}

\let\baraccent=\= 
\renewcommand{\=}[1]{\stackrel{#1}{=}} 

\usepackage{soul}

\usepackage[anticlockwise]{rotating}
\usepackage{float}
\usepackage{epstopdf}
\usepackage{gensymb}
\usepackage{enumitem}
\usepackage{xfrac}
\usepackage[nice]{nicefrac}
\usepackage[small]{titlesec}
\usepackage[toc,page]{appendix}
\usepackage[colorlinks,bookmarks=false,citecolor={black},linkcolor=blue,urlcolor=blue]{hyperref}
\usepackage{caption,todonotes,multirow,amsthm,array,makecell,bm}
\usepackage[backend=biber,style=ieee,citestyle=numeric-comp,sorting=none,natbib=true,isbn=false,url=false,eprint=false, hyperref=true,maxbibnames=1,maxcitenames=1,mincitenames=1,minbibnames=1]{biblatex} 
\usepackage{csquotes}
\DeclareSourcemap{
  \maps[datatype=bibtex, overwrite=true]{
    \map{
      \step[typesource=misc, typetarget=online]
    }
  }
}
\AtEveryBibitem{\clearfield{month}}
\DeclareFieldFormat[article]{volume}{
#1}
\DeclareFieldFormat[article]{pages}{
#1}
\renewbibmacro*{volume+number+eid}{%
  \printfield{volume}%
  \setunit*{\addnbspace}
  \printfield{number}%
  \setunit{\addcomma\space}%
  \printfield{eid}}
\DeclareFieldFormat[article]{number}{\mkbibparens{#1}}

\begin{document}
\title{Comparison of filament properties in real-size GBS simulations and experiments of TCV-X21}
\author{Y. Wang$^1$, C. Wüthrich$^{1}$, C. Theiler$^1$, S. García Herreros$^1$, D.S. Oliveira$^1$, D. Mancini$^1$, T. Golfinopoulos$^2$, P. Ricci$^1$, T. Body$^3$ and the TCV team$^4$}
\address{ $^1$ {\'E}cole Polytechnique F{\'e}d{\'e}rale de Lausanne (EPFL), Swiss Plasma Center (SPC), CH-1015 Lausanne, Switzerland,\\ 
$^2$ MIT Plasma Science and Fusion Center, MA-02139, Cambridge, USA,\\
$^3$ Commonwealth Fusion Systems, Devens, Massachusetts 01434, USA,\\
$^4$ See the author list of B.P. Duval et al 2024 Nucl. Fusion 64 112023
} 
\ead{yinghan.wang@epfl.ch}
\vspace{10pt}

\begin{abstract}
A direct quantitative comparison of Scrape-Off Layer (SOL) filament properties from fluid turbulence simulations using the GBS code and from experiments on the TCV tokamak is performed within the TCV-X21 validation case. This comparison is made possible by extending the open TCV-X21 dataset with 2D turbulence measurements obtained with Gas Puff Imaging (GPI), providing critical information on the size, velocity, and other key characteristics of turbulent filaments at the outboard midplane and in the divertor region. For the comparison, GBS simulations of TCV-X21 are analyzed using a dedicated synthetic GPI diagnostic that models the neutral helium-plasma interaction and emission processes and accounts for line-integration effects. Poloidal and radial filament velocities are found to be in good agreement between simulations and experiments, while the simulations overestimate the filament radial and poloidal sizes and underestimate the relative fluctuation levels. The simulations further indicate that filaments in the SOL are predominantly represented by density perturbations rather than temperature perturbations, consistent with previous assumptions in experimental analyses of cross-field turbulent transport from GPI data. The poloidal velocity direction of the filaments agrees with the time-averaged $\boldsymbol{E}\times\boldsymbol{B}$ direction at the outboard midplane and X-point region, but not in the divertor leg. Possible explanations are proposed and discussed, highlighting the influence of the instantaneous $\boldsymbol{E}\times\boldsymbol{B}$ velocity components in both poloidal and radial directions. This study provides new insights into turbulent filament behavior and contributes to guiding future efforts to improve first-principles simulations of the boundary plasma.

\nopagebreak
\end{abstract}

\nopagebreak
\ioptwocol
\section{Introduction}
\label{sec:introduction}
Understanding and predictive capabilities of tokamak boundary turbulent transport are needed for the successful operation of magnetic confinement fusion devices such as ITER and DEMO. Plasma turbulence in the edge and Scrape-off Layer (SOL) shapes the anomalous cross-field transport from the confined region with closed field lines towards the plasma facing components. This influences both the confinement performance of the core and the heat and particle load at the divertor and first wall of the device, which have to be mitigated~\cite{pittsPhysicsBasisFirst2019, zohmEUStrategySolving2021}. The importance of predicting these processes has further enhanced with the decision of ITER to use a full-tungsten wall~\cite{loarteNewITERBaseline2025,pittsPlasmawallInteractionImpact2025}. 

Among different types of turbulence features in the boundary plasma, distinct kinds of structures called 'blobs' are observed and identified in various magnetic confinement fusion devices~\cite{zwebenEdgeTurbulenceMeasurements2007}. These blobs, also known as 'filaments' as they are highly elongated along magnetic field lines, featuring locally enhanced density and temperature, are known to contribute significantly to the turbulence-induced cross-field particle and energy transport and hence gained significant attention~\cite{dippolitoConvectiveTransportIntermittent2011b,carraleroExperimentalValidationFilament2015a,theilerCrossFieldMotionPlasma2009, carraleroRoleFilamentsPerpendicular2018a,garciaFluctuationsTransportTCV2007,kirkLmodeFilamentCharacteristics2016}. 

 Substantial progress has been achieved in theory and numerical simulations toward understanding SOL turbulence and transport in general~\cite{stegmeirGlobalTurbulenceSimulations2019a,giacominFirstPrinciplesDensityLimit2022,zholobenkoTokamakEdgeSOLTurbulence2024}, and filaments in particular~\cite{russellBlobDynamics3D2004a, myraCollisionalityMagneticGeometry2006b,parutaBlobVelocityScaling2019}. Code validation via comparison of experiments and realistic, full-size, nonlinear turbulence simulations, is an essential step to gain insights in the turbulence transport phenomena and improve the numerical models and their predictive capabilities~\cite{ricciMethodologyTurbulenceCode2011a}. Recently, at the Tokamak à Configuration Variable (TCV)~\cite{duvalExperimentalResearchTCV2024}, a specially designed experimental reference case, TCV-X21, has been introduced as an optimized scenario for the validation of edge turbulence simulations in diverted geometry~\cite{oliveiraValidationEdgeTurbulence2022}. This scenario features a lower single-null, sheath-limited plasma. A publicly available experimental dataset with extensive measurements from a large set of diagnostics has been employed for the validation of transport codes~\cite{wangValidationSOLPSITERSimulations2024}, fluid turbulence codes~\cite{oliveiraValidationEdgeTurbulence2022,dudsonValidationHermes3Turbulence2025}, and gyrokinetic codes ~\cite{ulblInfluenceCollisionsValidation2023} in this scenario. However, up to now, observables related to 2-D filament structures in the poloidal cross-section have not been included in this dataset. This is a critical missing element, considering the impact these structures have on plasma interaction with the first wall, as well as the significant difference in the predictions of the structures from the different codes in this regard in ~\cite{oliveiraValidationEdgeTurbulence2022} (See figure 4 in this reference).

Experimentally, a key diagnostic on TCV to obtain SOL turbulence information is the Gas Puff Imaging (GPI) system~\cite{offedduGasPuffImaging2022a}, which provides 2-D turbulence observables at different locations of the SOL with high time resolution. GPI diagnostics have been used widely in multiple magnetic confinement fusion devices and have provided key information, including filament velocity scalings, statistical features, and quantification of their contribution to the cross-field transport~\cite{zwebenInvitedReviewArticle2017}. In TCV, L-mode studies with GPI provided new insights on the filament's 3-D characteristics~\cite{offedduCrossfieldParallelDynamics2022} and their properties in the X-point region~\cite{wuthrichXpointDivertorFilament2022a}. As such, GPI is a suitable tool for a direct comparison with full-size turbulence simulations. 

This paper contributes to the understanding of filaments and to code validation by presenting a comparison of 2-D filament properties between simulations and experiments on the TCV tokamak. A  synthetic diagnostic is developed and applied to the GBS simulations of the TCV-X21 scenario~\cite{giacominGBSCodeSelfconsistent2022a,oliveiraValidationEdgeTurbulence2022,offedduGasPuffImaging2022a}, followed by direct qualitative and quantitative comparisons using the analysis techniques developed for filaments in TCV. This work further extends the TCV-X21 dataset, allowing more stringent model validation. 

This paper is organized as follows. After the introduction in section~\ref{sec:introduction}, the experiment, diagnostics, and GBS simulation setup are described in section~\ref{sec:model_experiment}. Next, we show the workflow and working principle of the synthetic GPI diagnostics in section~\ref{sec:synthetic_GPI_diagnostic}. Then, we present the qualitative and quantitative comparison results between simulations and experiments in section~\ref{sec:qualitative_comparison} and section~\ref{sec:quantitative_filament_comparison}, respectively. A deeper investigation of the mechanism of the poloidal velocity of the filaments in different SOL regions is given in section~\ref{sec:velocity_contribution}. After a discussion in section~\ref{sec:discussion}, we conclude the paper in section~\ref{sec:conclusion}.

\section{Experimental scenarios and simulations} \label{sec:model_experiment}
\subsection{Experiments and GPI measurements}
\begin{figure*}[htbp]
\includegraphics[width= 1.0\textwidth ]{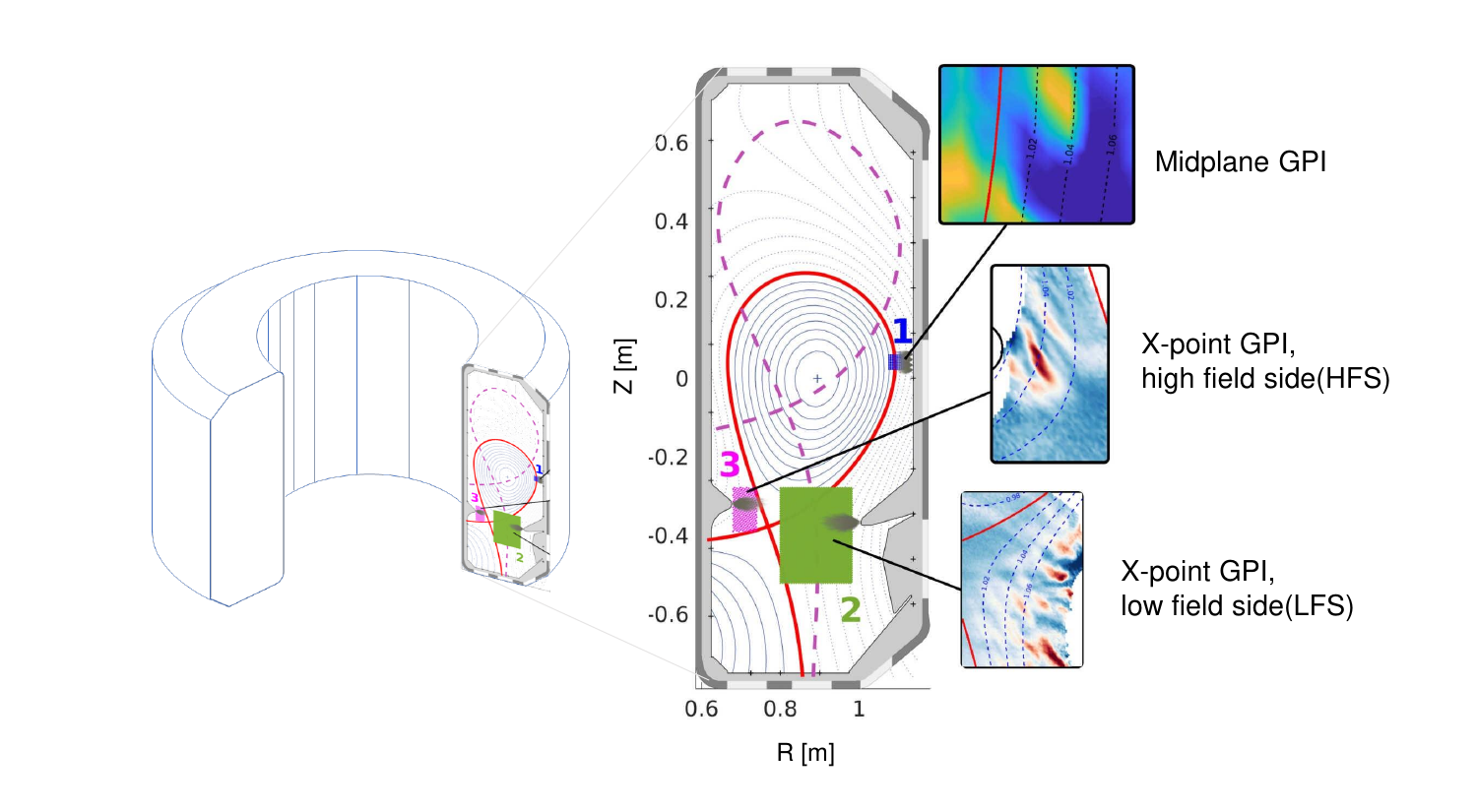}
\caption{ \label{fig:GPI_diag}
\textbf{GPI diagnostics in TCV.} The locations and fields of view of the midplane GPI and the Xpt GPI configurations are shown in the same poloidal cross-section, together with some typical 2-D plasma fluctuation snapshots. Examples of plasma equilibrium flux surfaces and separatrices are plotted. The plasma can be positioned for X-point GPI measurements around the X-point  (red solid line) or around the divertor leg region (purple dashed line). Right image adopted from ~\cite{offedduGasPuffImaging2022a}.}
\end{figure*}

For the comparison, dedicated TCV-X21 plasma discharges (lower single null, low magnetic field, and low density~\cite{oliveiraValidationEdgeTurbulence2022}) have been repeated multiple times in reversed magnetic field direction to obtain GPI measurements at different locations, including the outboard midplane, X-point region, and the divertor leg region. In order to probe the divertor leg region, the plasma is carefully shifted upwards to be compatible with the Xpt GPI system, as shown in figure~\ref{fig:GPI_diag}. Equivalently~\cite{wuthrichDependenceDivertorTurbulence2025a}, the field of view is approximated to be shifted downwards, as marked by the blue dashed line in figure~\ref{fig:qualitative_comparison}. The discharges used in the analysis of this paper are summarized in table~\ref{tab:shot_summary} and the geometry is shown in figure~\ref{fig:qualitative_comparison}. 

For the experimental part, the GPI diagnostics in TCV (figure~\ref{fig:GPI_diag}) ~\cite{offedduGasPuffImaging2022a} utilize actively injected neutral gas (here helium) to enhance the emission from plasma-neutral interaction, thus revealing the patterns and dynamics of the plasma fluctuations at a high spatiotemporal resolution. For the outboard midplane GPI system, the field of view covers approximately $5\times 4 \ \mathrm{cm^2}$, using $12 \times 10$ channels of optical fibers, connected to Avalanche Photon Diode (APD) sensors. For the X-point GPI system, the brightness information is captured by a Phantom v2012 high-speed camera with $96\times 128$ pixels covering a field of view of approximately $17.5\times 22 \ \mathrm{cm^2} $.

\begin{table*}[]
\centering
    \caption{ \label{tab:shot_summary}\textbf{Summary of discharges in this work with featuring GPI measurements.} } 
    \begin{tabular}{llllll}
    \hline
    discharge & Field direction & Baffles  & GPI gas \& gas flux & GPI position & GPI observation time \\ \hline
    70336          &  Reversed           & SILO &   He  \ $6\times 10^{19}\ \mathrm{atom/s}$                &    X-point          &   $1.52-1.62 \ \mathrm{[s]}$                   \\
    70545          &  Reversed           & SILO &     He \ $6\times 10^{19}\ \mathrm{atom/s}$               &    Divertor leg          &   $1.53-1.63 \ \mathrm{[s]}$                   \\
    77027          &  Reversed           & none &     He \ $2\times 10^{19}\ \mathrm{atom/s}$                &   Outboard midplane           &  $1.40-1.50 \ \mathrm{[s]}$                    \\ \hline
    \end{tabular}
\end{table*}

\subsection{The GBS simulation}
For the comparison in this study, we use numerical simulation results from the GBS code ~\cite{ricciSimulationPlasmaTurbulence2012a,giacominGBSCodeSelfconsistent2022a}. For the present simulation, GBS solved 3D fluid Branginskii equations with assumed ionization locations of neutrals located in outer  core region. The spatial resolution in the poloidal plane was $\sim 2 \ \mathrm{mm}$. In the toroidal direction, the resolution is $\sim 2.8 \degree$. The detailed simulation setup can be found in Ref.\cite{oliveiraValidationEdgeTurbulence2022}. 

In this work, we applied our analysis to the reversed-field, TCV-X21 simulation output during the steady fluctuation phase starting at $0.347 \ \mathrm{ms}$, with a time resolution of $0.002 \ \mathrm{ms}$.

\section{Synthetic GPI diagnostics} 
\label{sec:synthetic_GPI_diagnostic}
In order to conduct direct comparisons between simulations and experiments, in this work, a synthetic GPI diagnostic is developed and adapted to post-process the GBS outputs. 

The synthetic diagnostic is constructed by extending the 0-T model, which was developed for the design process of the X-point GPI diagnostic in TCV~\cite{offedduGasPuffImaging2022a}. The model was originally used for estimating the expected signal level for the X-point GPI system, given a time-averaged plasma background density and temperature. Here, we extended this model such that it can post-process the outputs from numerical simulations frame-by-frame and compute the brightness structures dynamically in the same format and at the same position as in the experiments.

\begin{figure}[htbp]
\includegraphics[width= 0.48\textwidth ]{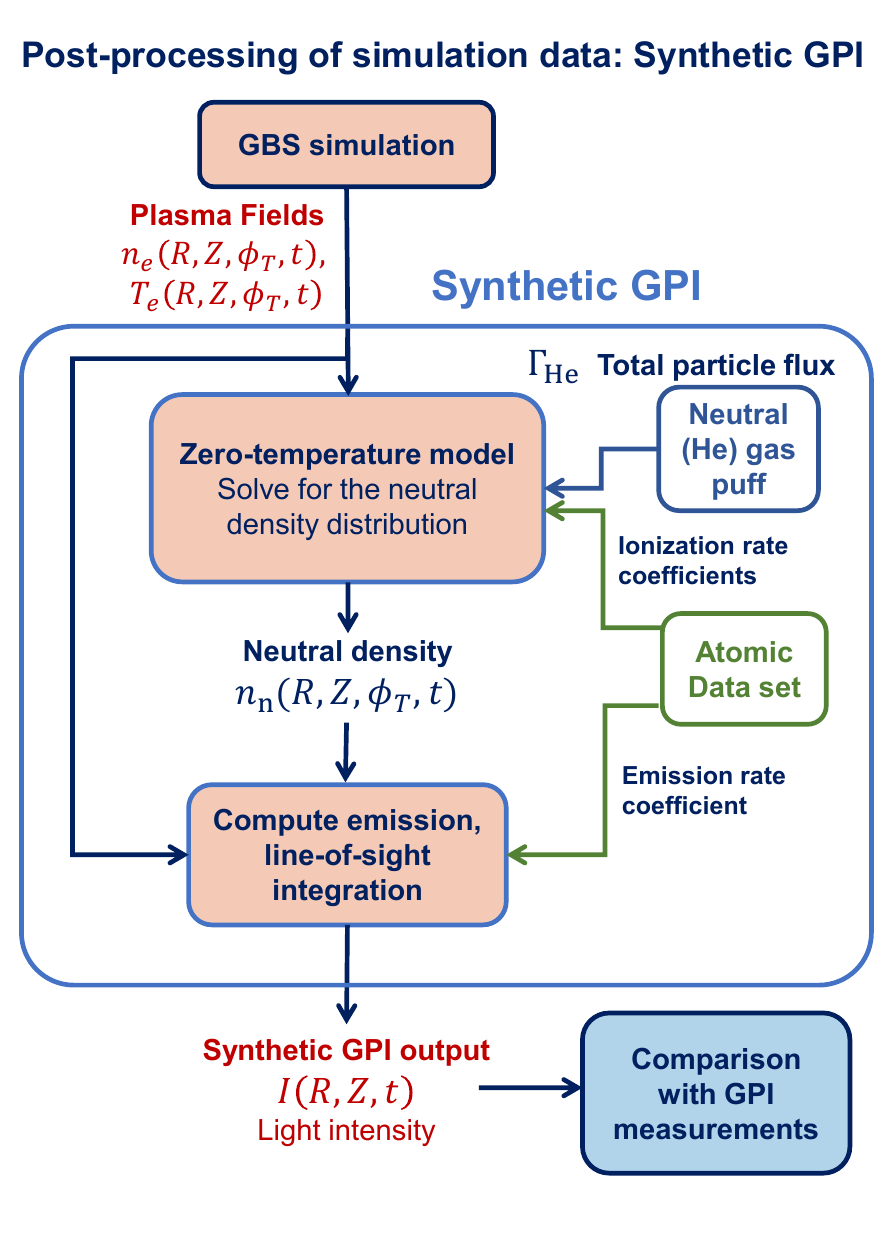}
\caption{ \label{fig:synGPI_flowchart}
\textbf{Workflow of the synthetic GPI diagnostic.}}
\end{figure}

\begin{figure*}[htbp]
\includegraphics[width= 1.0\textwidth ]{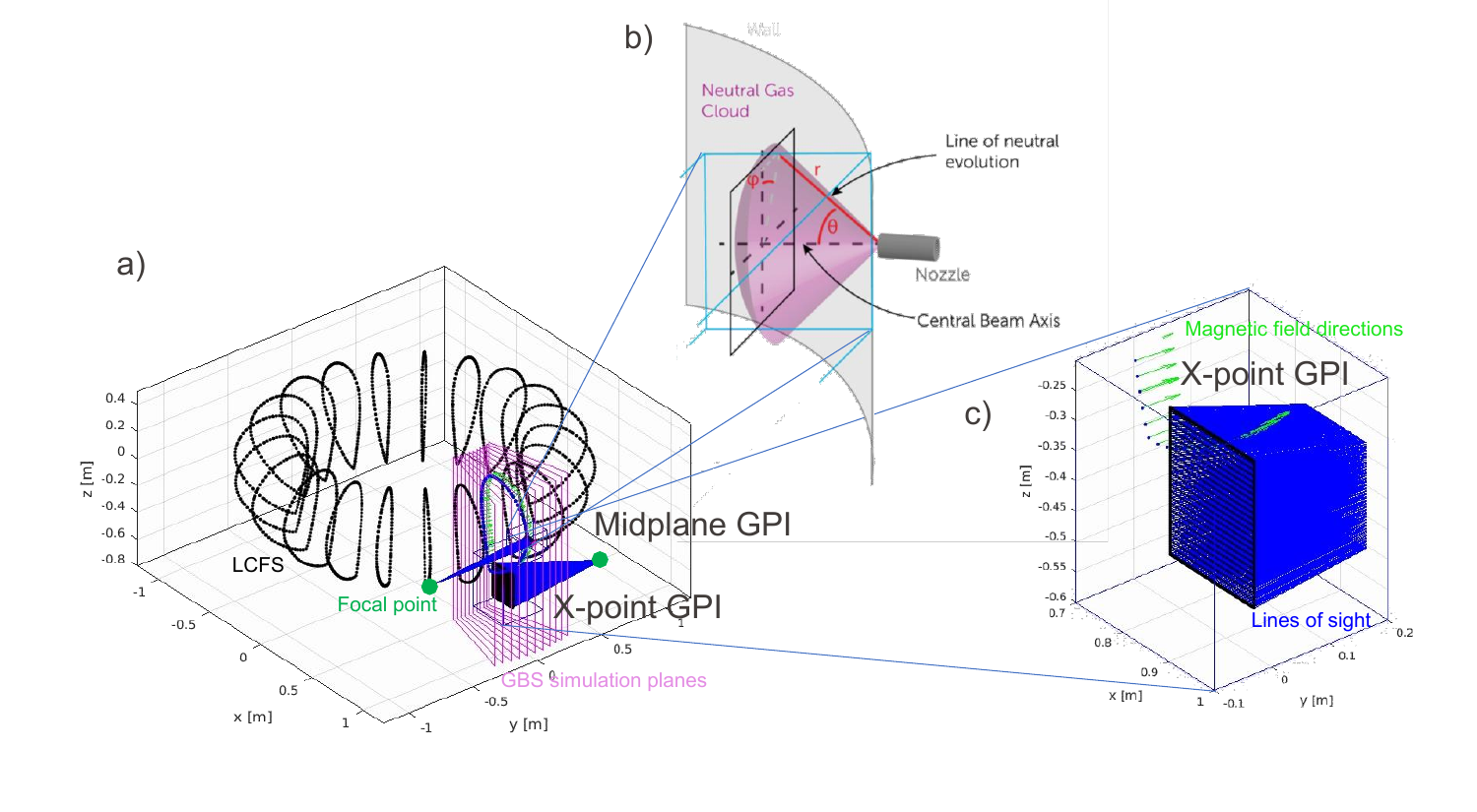}
\caption{ \label{fig:synGPI}
\textbf{Illustration of the geometry and working principle of the synthetic GPI diagnostic.} (a) The LCFS at multiple toroidal angles are shown in dark dotted lines. The magnetic field line at one poloidal plane is plotted as green vectors (also shown in c) ). Multiple toroidal planes of simulation results (purple planes) from the turbulence simulation (here GBS) are provided as an input. The GPI focal point (green circle) and the lines of sight (blue lines) are shown for both midplane GPI and X-point GPI. (b)The simulation domain of the neutral gas injection and reaction with the plasma, together with its spherical coordinate system. (c)A zoomed-in view of the field of view and lines of sight of the X-point GPI system. The magnetic field direction is plotted in green. For visualization, only part of the $96\times 128$ lines of sight are plotted.}
\end{figure*}

As shown in  figures~\ref{fig:synGPI_flowchart} and ~\ref{fig:synGPI}, using 3-D background profiles of density and temperature provided by a simulation, the synthetic diagnostic models the streaming and ionization loss of the injected neutral gas by solving a continuity equation in a spherical coordinate system (figure~\ref{fig:synGPI} (b)). This yields the resulting density distribution of the neutral particles. Then, using the emission rate coefficients provided by an atomic database (here~\cite{summersAtomicDataModelling2011}), the model calculates the emissivity of the neutral particles in the simulated domain. Finally, via integration along the lines of sight of the GPI system, the model provides 2-D brightness information in the same format as from experimental measurements. 

In this work, compared to the 0-T model in ~\cite{offedduGasPuffImaging2022a}, the realistic geometry of the GPI lines of sight and the 3-D GBS input geometry are considered, as shown in figure~\ref{fig:synGPI}. The gas injection location is set up as in realistic cases to obtain different fields of view, as shown in figure~\ref{fig:qualitative_comparison}. Under the frozen plasma assumption, this synthetic diagnostic can be applied to the GBS density and temperature output frame by frame. Specifically, the frozen plasma assumption means that, during one time frame of the neutral distribution calculation, the plasma is static. In this model, we also assume that the helium neutrals move along straight trajectories before ionizing, without considering momentum exchanges due to collisions with ions or neutrals.  This model further does not consider the feedback of the injected neutrals on the plasma.

\section{Qualitative comparison of brightness
fluctuation snapshots} \label{sec:qualitative_comparison}
In this section, we present an example of the synthetic diagnostic output and a qualitative comparison of the turbulence and filament properties. The synthetic diagnostic itself also provides useful insights on the relationship between fluctuations of the GPI brightness and the plasma physical quantities (density and temperature), as also discussed.

\begin{figure*}[htbp]
\includegraphics[width= 1.0\textwidth ]{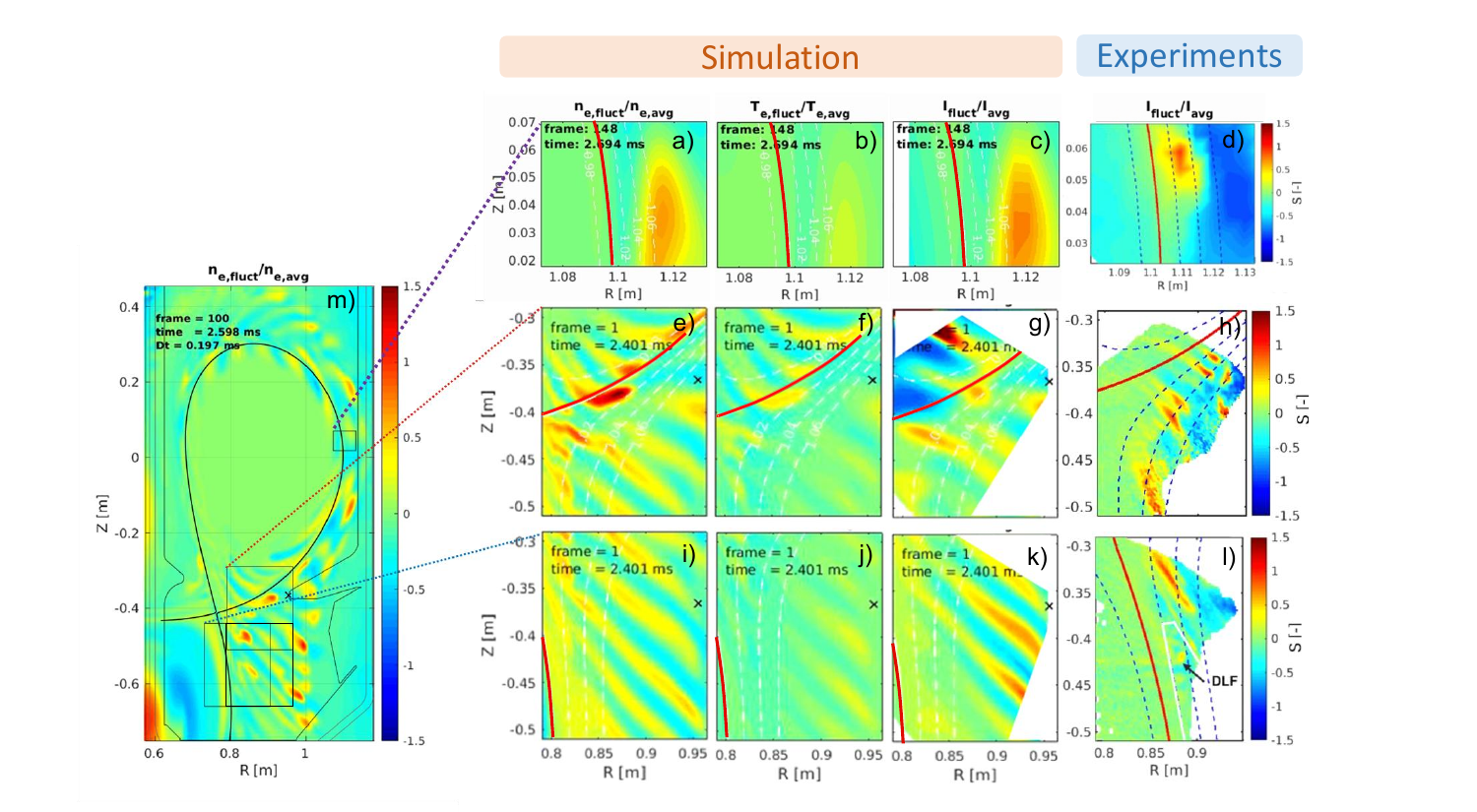}
\caption{ \label{fig:qualitative_comparison}
\textbf{Qualitative comparison of the turbulence patterns at the different poloidal locations.} Snapshots of the relative fluctuation of electron density $n_e$, electron temperature $T_e$, and brightness from simulations and experiments are plotted, respectively.}
\end{figure*}

In figure.~\ref{fig:qualitative_comparison}, a set of 2D snapshots is displayed. They show the relative fluctuation pattern of electron density $n_e$, electron temperature $T_e$, and brightness $I$ from the simulation, with the latter obtained from the synthetic diagnostic. For comparison, the last column shows snapshots of the relative fluctuation level of brightness from the experiment in the same scenario. 

The relative fluctuation $\Tilde{a}$ of a quantity $a$ is defined here as the quantity at a certain time frame $a(t)$ with its time average $\langle a\rangle $ subtracted (the fluctuation part), normalized by the time average $\langle a \rangle$. 
\begin{equation}
\label{eq:relative_fluct}
\Tilde{a} = \dfrac{a_{\text{fluct}}}{a_{\text{avg}}}=\dfrac{a-\langle a\rangle}{\langle a\rangle}
\end{equation}
In the experiments, a moving $1\ \mathrm{ms}$ average is conducted to provide the average value, following the same methods as in previous works~\cite{offedduCrossfieldParallelDynamics2022,wuthrichXpointDivertorFilament2022a}. In the simulation, the data is averaged over the whole time window of the simulation of $0.35 \ \mathrm{ms}$. 

Comparing the first and the second column of figure~\ref{fig:qualitative_comparison}, we observe that there is a correlation between the relative fluctuation of density and temperature. Another clear feature is that the fluctuation magnitude of temperature appears lower than that of the density. This observation is consistent with the previous assumptions made in the GPI experimental analysis, where brightness fluctuations in the SOL were interpreted to be primarily due to density fluctuation~\cite{wuthrichXpointDivertorFilament2022a}. Then, comparing the first and the third column, the relative fluctuation of the brightness is shown to be correlated with the relative fluctuation of the density outside the separatrix. However, inside the separatrix, this correlation no longer holds. This could be related to the neutral gas depletion when it gets ionized in the high-temperature core region. In the study of filaments in the SOL (outside the separatrix), this simulation, however, supports the assumptions that the brightness fluctuation is strongly correlated with the density fluctuation~\cite{coroadoNumericalSimulationsGas2022a}.  

Next, by comparing the relative fluctuations of brightness between the simulation and experiments, some qualitative observations can be obtained. In this single snapshot, the size of the simulated filaments at the outboard midplane is larger than the experimental case. In the X-point and the divertor leg regions, in both simulation and experiments, we observe elongated structures with enhanced brightness compared to the average. The simulations reproduce the stretched shape in the direction normal to the flux surface. This is consistent with the interpretation in experiments that due to the flux expansion, the upstream connected filaments appear highly elongated in the X-point and divertor region~\cite{wuthrichXpointDivertorFilament2022a}. Besides the similarities, the comparison also reveals some mismatches. First, the sizes in the X-point and divertor region are overestimated in the simulation. Secondly, comparing the simulation and the experiments, the number of individual filament structures is lower in the simulation than in the experiments, which is also partially related to the overestimated sizes of the filaments. Thirdly, the peak magnitude of the fluctuation in the simulation is somewhat lower than that in the experiments. From the experimental data in the divertor leg region (figure~\ref{fig:qualitative_comparison} l), we can observe divertor localized filaments (DLFs)~\cite{wuthrichXpointDivertorFilament2022a,wuthrichDependenceDivertorTurbulence2025a}, featuring a round structure in contrast to the filed aligned, upstream-connected structures. In the simulation, these structures can not clearly be identified in the current data. 

It is worth noting that the qualitative analysis presented in this section is based only on a single, yet representative, frame. To further confirm these observations and study the dynamics of the filaments, we conduct a quantitative analysis using data from the whole time range of the simulation dataset, presented in the next section.

\section{Quantitative simulation-experiment comparison of filament properties} \label{sec:quantitative_filament_comparison}
\begin{figure*}[htbp]
\includegraphics[width= 1.0\textwidth ]{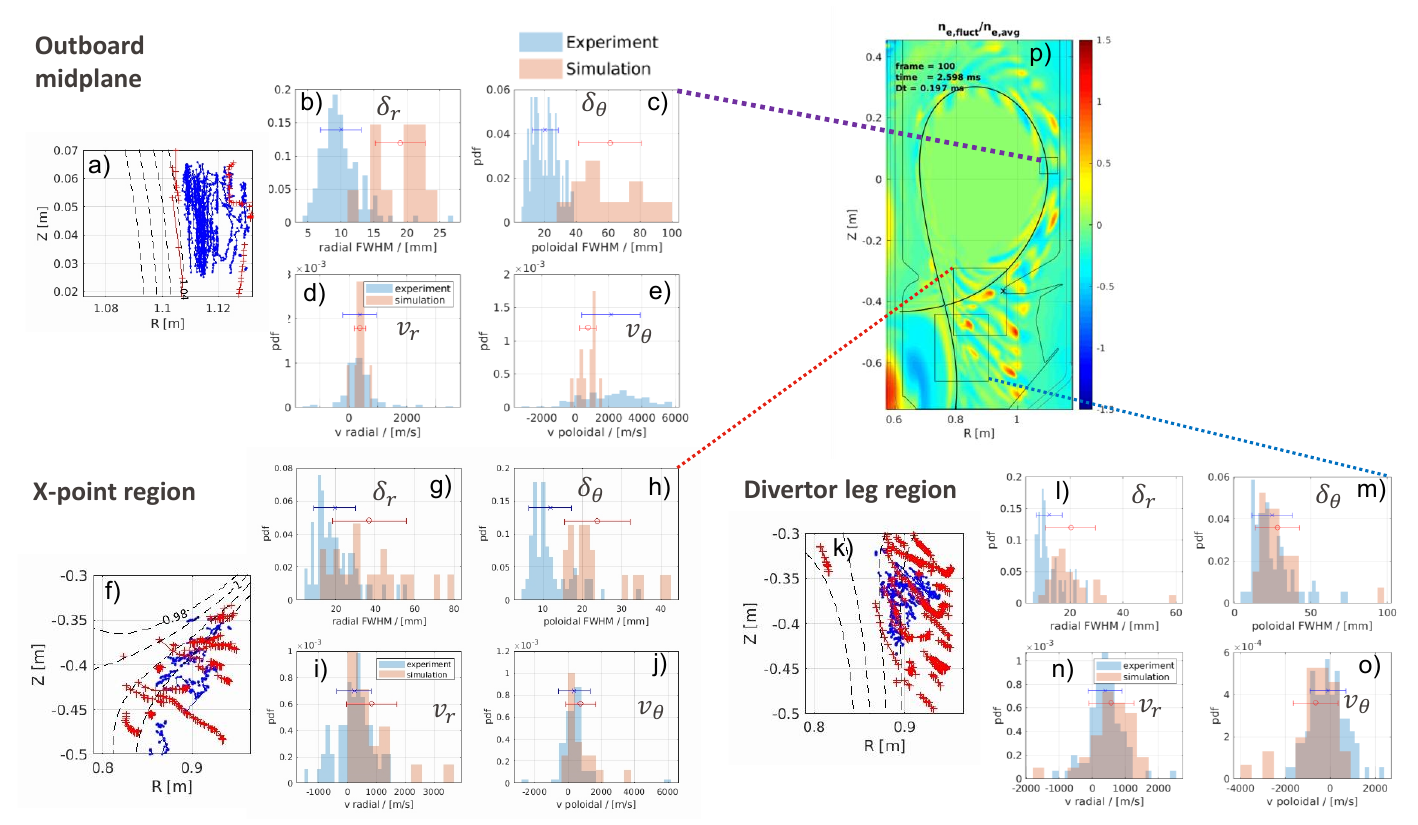}
\caption{ \label{fig:tracking_pdf_Rev}
\textbf{The filament tracking results at three locations for both the simulation and experimental data.} The typical trajectories of the filaments are plotted in a), f) and k), with experimental trajectories in blue and simulated trajectories in red. In the other subplots, the probability density function (PDF) of the filament sizes ($\delta_r$ and $\delta_\theta$) and velocities ($v_r$ and $v_\theta$) in the radial (subscript $r$) and poloidal directions (subscript $\theta$) are plotted. The simulation data is plotted in orange and the experimental data is plotted in blue. The average value and the standard deviation of the distributions are shown by markers and bars, respectively. }
\end{figure*}

In this section, we analyze the filament properties in a quantitative way. Due to the limited number of events from the GBS simulation, we apply the individual blob (filament) tracking algorithm as described in Ref.~\cite{offedduAnalysisTechniquesBlob2023} and study the statistical distribution of the properties of the filaments. We will mainly focus on the sizes and velocities of the filaments, which are key factors influencing the cross-field transport. 

In figure~\ref{fig:tracking_pdf_Rev}, at the three locations included in this study, we illustrate the typical trajectories of the blobs captured in the simulations (red lines) and in the experiments (blue lines), as well as the probability density function (PDF) of the blob sizes ($\delta_r$ and $\delta_\theta$) and velocities ($v_r$ and $v_\theta$) in the radial (subscript $r$) and poloidal directions (subscript $\theta$). The average value and the standard deviation of the distributions is shown by markers and bars, respectively. 

At the outboard midplane (figure~\ref{fig:tracking_pdf_Rev} (a-e)), to improve the statistics of the filament analysis in the simulation data, the synthetic diagnostic is applied at two toroidally separated locations in the simulation domain, $180^\circ$ apart. The events captured show that the simulated filament structures are larger than those in the experiments. The overestimation factor is approximately 2 in the radial direction and approximately 3 in the poloidal direction. The velocity in the radial and poloidal direction is, however, well reproduced by the simulation, with the mean simulated value lying within the standard deviation. Due to the relatively limited number of events captured by the tracking algorithm, the distribution shape is not fully converged in the statistical analysis of the simulation. In the experiments, filaments appear to have a larger population at poloidal velocities above $2\ \mathrm{km/s}$.

In the X-point region and the divertor leg region, the experimental velocity distribution is also well reproduced by the simulation. The simulated average velocity matches the experimental value within its standard deviation. The simulation still overestimates the filament radial and poloidal size by a factor of $2$ in the X-point region. However, in the divertor leg region, the averaged poloidal size matches the experiment.  In both regions, the location and orientation of the trajectories of filaments show a good match.

\section{Study of the mechanism of the filament poloidal velocity}\label{sec:velocity_contribution}
A key agreement found in the quantitative comparisons described above is that the velocities in the poloidal and radial directions are generally well reproduced by the simulation. This enables a deeper investigation on the physical mechanism of the filament motion. 

\begin{figure*}[htbp]
\centering
\includegraphics[width= 0.8\textwidth ]{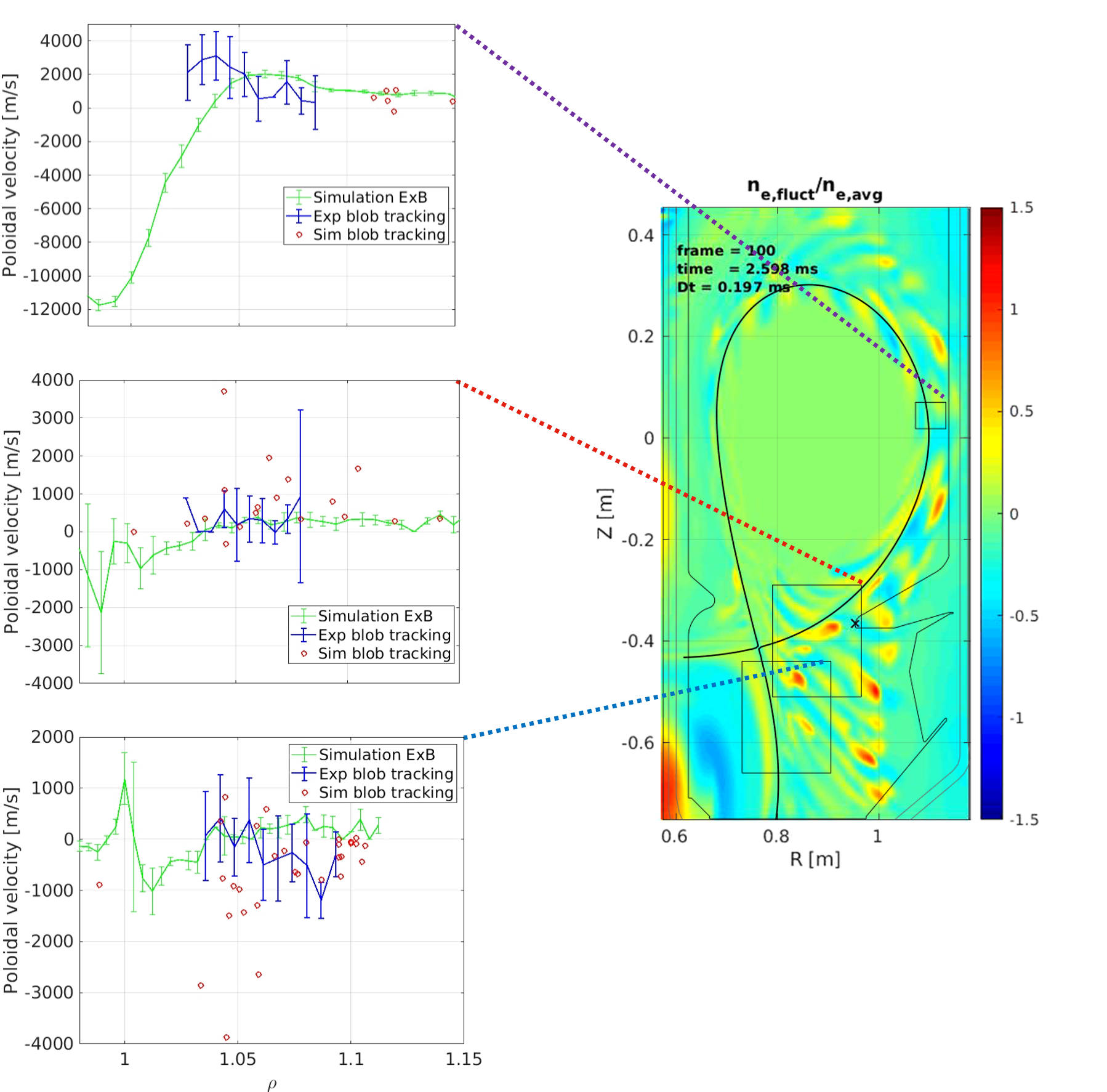}
\caption{ \label{fig:velocity_ExB}
\textbf{The filament poloidal velocity profile from simulation and experiment compared with the simulated, time-averaged poloidal $\boldsymbol{E}\times \boldsymbol{B}$ velocity profile in different regions as indicated.}}
\end{figure*}

The poloidal motion of the filaments in the SOL is believed to be mainly determined by the averaged $\boldsymbol{E}\times \boldsymbol{B}$ drift. Therefore, in figure~\ref{fig:velocity_ExB}, we plotted the poloidal component of the filament velocity versus the flux coordinate $\rho_\psi$ for the three locations studied in this work. For the experimental result, the filament events are binned in uniform $\rho_\psi$ intervals, and the mean velocity values are plotted in blue, with the error bars indicating the standard deviation of the data in each bin. The filament events tracked in the simulations are indicated by red circles. Then, from the simulations, the time-averaged poloidal $\boldsymbol{E}\times \boldsymbol{B}$ drift velocity is calculated. The data are binned within intervals of $\rho_\psi$, resulting in the green curve with error bars showing the standard deviation. 

One can observe that for midplane filaments, the simulated filament poloidal velocities match well with the averaged poloidal $\boldsymbol{E}\times \boldsymbol{B}$ drift velocity. Fair agreement is also found for the experimental filament velocities and the calculated poloidal $\boldsymbol{E}\times \boldsymbol{B}$ velocity for $\rho_\psi>1.05$.

In the X-point region (figure~\ref{fig:velocity_ExB} (b)), we also observe that most of the simulated filaments have a poloidal velocity consistent with the averaged poloidal $\boldsymbol{E}\times \boldsymbol{B}$ drift velocity. The experimental values are also consistent with the simulation curve. Consistently, in these two regions, we find for these reversed field plasmas a positive, i.e. counter clockwise poloidal velocity.

In the divertor leg region, instead, we find different results compared with the previous two cases. The filament tracking velocities in the experiments and simulations still match well with each other. However, they no longer agree with the local, time-averaged poloidal $\boldsymbol{E}\times \boldsymbol{B}$ drift velocity, especially for $\rho_\psi>1.05$. They actually have the opposite sign. The averaged poloidal $\boldsymbol{E}\times \boldsymbol{B}$ velocity is still positive, while the tracked filament velocities are negative. This simulation finding aligns well with the experimental observation in \cite{wuthrichXpointDivertorFilament2022a}. A further investigation of this observation is given in Section~\ref{sec:discussion}.

\section{Discussion}\label{sec:discussion}
The experiment-simulation comparison of filament properties in this reversed-field TCV-X21 scenario generally shows that the simulation overestimates the filament size while capturing well the velocity. This observation provides useful directions for the improvement of the turbulence simulation. Possible reasons for this overestimation include: The lack of divertor neutrals used in the GBS simulation, the Boussinesq approximation applied in the present GBS simulation, and the resistivity used in the GBS simulation, which, due to numerical considerations, is higher than the experimental values~\cite{oliveiraValidationEdgeTurbulence2022}. Simulations with the above aspects improved, partially guided by the present findings, are the subject of ongoing work.

\begin{figure}[htbp]
\centering
\includegraphics[width= 0.45\textwidth ]{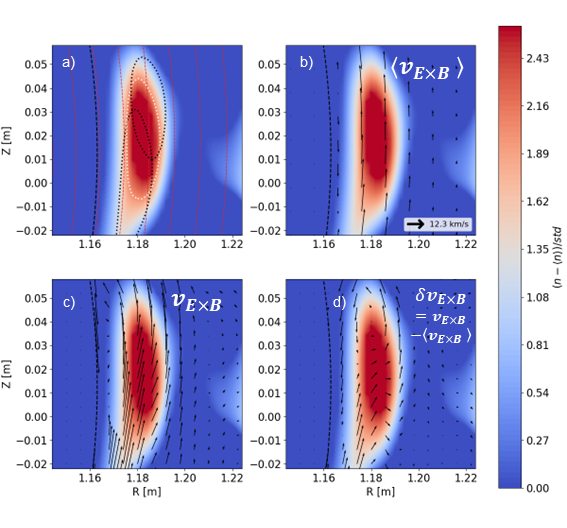}
\caption{ \label{fig:ExB_quiver_midplane}
\textbf{Outboard midplane 2D snapshot showing the cross-field motion of a filament together with different components of the $\boldsymbol{E}\times \boldsymbol{B}$ drift.} The relative density fluctuation is shown by the colormap in a) to d), where the colormap is cropped to 0 for better visualization. In a), the red dotted contours show the flux surfaces, and the white dotted contour shows a filament event. The black dotted contours show the position $11\ \mu s$ before and $11\ \mu s$ after the current time frame. b) shows the averaged $\boldsymbol{E}\times \boldsymbol{B}$ velocity in black vectors. c) shows the total instantaneous $\boldsymbol{E}\times \boldsymbol{B}$ velocity. d) shows the instantaneous fluctuation of the $\boldsymbol{E}\times \boldsymbol{B}$ velocity.}
\end{figure}

\begin{figure}[htbp]
\centering
\includegraphics[width= 0.5\textwidth ]{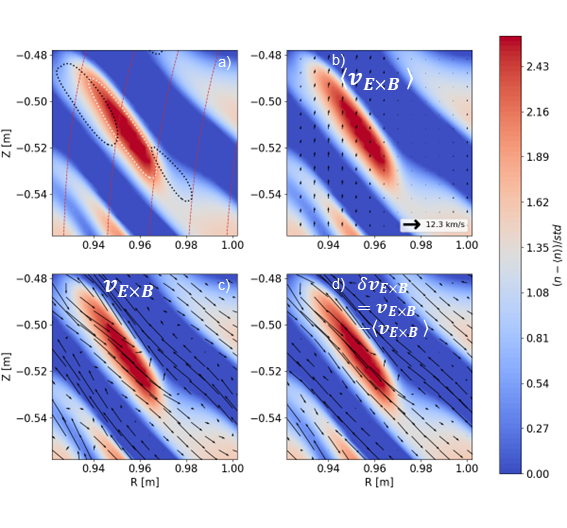}
\caption{ \label{fig:ExB_quiver_divertor}
\textbf{Divertor leg 2D snapshot showing the cross-field motion of a filament together with different components of the $\boldsymbol{E}\times \boldsymbol{B}$ drift.} The relative density fluctuation is shown by the colormap in a) to d), where the colormap is cropped to 0 for better visualization. In a), the red dotted contours show the flux surfaces, and the white dotted contour shows a filament event. The black dotted contours show the position $11\ \mu s$ before and $15.4\ \mu s$ after the current time frame. b) shows the averaged $\boldsymbol{E}\times \boldsymbol{B}$ velocity in black vectors. c) shows the instantaneous total $\boldsymbol{E}\times \boldsymbol{B}$ velocity. d) shows the instantaneous fluctuation of the $\boldsymbol{E}\times \boldsymbol{B}$ velocity.}
\end{figure}

The reason for the difference between the poloidal velocity of filaments and the time-averaged, poloidal  $\boldsymbol{E}\times \boldsymbol{B}$ velocity in the divertor leg region, seen in experiment and simulation in Sec.~\ref{sec:velocity_contribution}, is further investigated and discussed here. We will focus on two possible driving mechanisms for the poloidal motion of filament structure in the divertor region, assuming that they extend as a coherent structure along the magnetic field up to the outboard midplane. The first mechanism is the local, instantaneous electric field that drives the $\boldsymbol{E}\times \boldsymbol{B}$ drift. The second mechanism we consider is described as follows: The upstream connected filaments, as a whole, feature a much faster parallel transport compared to the perpendicular transport. Therefore, the motion of the downstream (divertor leg region) pair of a filament could be determined by the motion of the filament at the outboard midplane, combined with the free streaming of the same filament particles along the magnetic field into the divertor. To investigate this, a deeper look at the cross-field motion of the simulated filaments is conducted in figure~\ref{fig:ExB_quiver_midplane} and figure~\ref{fig:ExB_quiver_divertor}, where the time-averaged $\boldsymbol{E}\times \boldsymbol{B}$ velocity, $\langle \vec{v}_{\boldsymbol{E}\times \boldsymbol{B}}\rangle$, the instantaneous $\boldsymbol{E}\times \boldsymbol{B}$ velocities, $\vec{v}_{\boldsymbol{E}\times \boldsymbol{B}}$, and the difference between the two (i.e. the fluctuation component), $\delta \vec{v}_{\boldsymbol{E}\times \boldsymbol{B}}$, are plotted over the snapshots of a blob motion instance. The figures show clear distinctions between the midplane case (figure~\ref{fig:ExB_quiver_midplane}) and the divertor leg case (figure~\ref{fig:ExB_quiver_divertor}). At the midplane, the averaged $\boldsymbol{E}\times \boldsymbol{B}$ velocity dominates the poloidal motion of the filament. In contrast, in the divertor leg region, the magnitude of the (upwards pointing) averaged $\boldsymbol{E}\times \boldsymbol{B}$ velocity is much smaller than the fluctuation part. The latter points to the bottom right, consistent with the filament motion. This is consistent with the findings in figure~\ref{fig:velocity_ExB}. Therefore, the simulation suggests that the poloidal motion of the filaments in the divertor leg region, like the radial motion, is dominated by the instantaneous $\boldsymbol{E}\times \boldsymbol{B}$, rather than by its time-averaged value.

\begin{figure*}[htbp]
\centering
\includegraphics[width= 0.9\textwidth ]{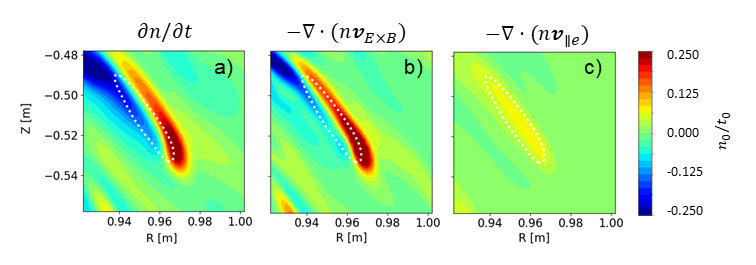}
\caption{ \label{fig:fig_divFlux}
\textbf{Snapshot from GBS simulation, showing three key terms from the density continuity equation.} a) the partial time derivative of density $\partial n/\partial t$, b) the negative divergence of the flux due to the $\boldsymbol{E}\times \boldsymbol{B}$ drift $-\nabla\cdot (n\boldsymbol{v}_{E\times B})$, and c) the negative divergence of the flux due to the parallel flow $-\nabla\cdot (n\boldsymbol{v}_{\parallel e})$.}
\end{figure*}

To further confirm the $\boldsymbol{E}\times \boldsymbol{B}$ drift as the cause of the filament movement in the divertor, as opposed to effects due to free streaming along the magnetic field from upstream, we plotted three key terms from the density continuity equation of the filaments in figure~\ref{fig:fig_divFlux}: the partial time derivative of the density, the divergence of the flux due to the $\boldsymbol{E}\times \boldsymbol{B}$ velocity and the divergence of the flux due to the parallel velocity. It clearly shows that the flux due to the $\boldsymbol{E}\times \boldsymbol{B}$ drift accounts for most of the temporal variation in density, confirming that the movement of the filaments is indeed due to cross-field, rather than parallel transport. 

The results presented above show the capability of the fluid model implemented in GBS to simulate the filament motion, despite the fact that these turbulence simulations still have some clear discrepancies compared to the experiments in terms of the plasma profiles~\cite{oliveiraValidationEdgeTurbulence2022}. The match found between the poloidal filament velocity and the time-averaged poloidal $\boldsymbol{E}\times \boldsymbol{B}$ drift velocity at the outboard midplane and X-point supports to use GPI to calculate the radial electric field in these regions, similarly as Doppler backscattering (DBS) is being used for this purpose inside the separatrix~\cite{rienackerSurveyEdgeRadial2025c}.  

\section{Conclusion}\label{sec:conclusion}
In this work, the 2-D filament properties in the TCV-X21 scenario on TCV were compared between full-size 3-D turbulence simulations and experiments. To enable this comparison, a synthetic GPI diagnostic was introduced and applied to the simulation outputs. Poloidal and radial filament velocities were found to be in good agreement between simulations and experiments. Compared to the experiments, the simulations overestimated, however, filament sizes (by a factor of 2-3) in radial and poloidal dimensions. In the simulation, filaments were predominantly associated with a density fluctuation and showed lower temperature fluctuations, which is consistent with previous assumptions in experimental analysis of cross-field turbulent transport from GPI data. Poloidal filament velocities were found not to follow the direction of the time-averaged poloidal $\boldsymbol{E}\times \boldsymbol{B}$ drift velocity in the divertor leg region, while they do follow it at the outboard midplane. Analysis of the simulation shows that the instantaneous $\boldsymbol{E}\times \boldsymbol{B}$ drift, instead of its mean value, contributes the most to the poloidal (and radial) filament motion in the divertor leg, and the cause of the filament motion in the poloidal plane is clearly the local cross-field transport, rather than parallel streaming of particle from a different region. 

On the path towards fully predictive simulations, a better simulation-experiment agreement of filament sizes will be needed. Several paths are currently being pursued: self-consistent inclusion of neutrals, removal of the Boussinesq approximation, and code improvement to allow for more realistic resistivity. 

The public TCV-X21 dataset has been extended here by the presented GPI data for further validations.

\section*{Acknowledgements}
This work was supported in part by the Swiss National Science Foundation. This work has been carried out within the framework of the EUROfusion Consortium, via the Euratom Research and Training Programme (Grant Agreement No 101052200 — EUROfusion) and funded by the Swiss State Secretariat for Education, Research and Innovation (SERI). Views and opinions expressed are however those of the author(s) only and do not necessarily reflect those of the European Union, the European Commission, or SERI. Neither the European Union nor the European Commission nor SERI can be held responsible for them.
\printbibliography[heading=bibintoc]

\end{document}